\pdfoutput=1
\documentclass{elsart}
\usepackage{graphicx,color}
\usepackage{amsmath,amssymb}
\usepackage{slashbox}
\usepackage{epstopdf}

\def\anp#1#2#3{Annals Phys. #1 (#3) #2}
\def\arnps#1#2#3{Ann.\ Rev.\ Nucl.\ Part.\ Sci.\ #1 (#3) #2}

\def\appb#1#2#3{Acta Phys.\ Polon.\  B #1, (#3) #2}
\def\cmp#1#2#3{Comm. Math. Phys. #1 (#3) #2}
\def\cnpp#1#2#3{Comments Nucl.\Part.\Phys.\ #1 (#3) #2}
\def\cnpp#1#2#3{Comments Nucl. Part. Phys. #1 (#3) #2}

\def\ibid#1#2#3{{\it ibid.} #1 (#3) #2}
\def\ijma#1#2#3{Intl. Jour. Mod. Phys. A #1 (#3) #2}

\def\jhep#1#2#3{J. High Energy Phys. #2 (#3) #1}
\def\jkps#1#2#3{J. Korean Phys. Soc. #1 (#3) #2}

\def\jpg#1#2#3{Jour. Phys. G #1 (#3) #2}

\def\npa#1#2#3{Nucl. Phys. A #1 (#3) #2}
\def\npb#1#2#3{Nucl. Phys. B #1 (#3) #2}

\def\pan#1#2#3{Phys. Atom. Nucl. #1 (#3) #2}
\def\plb#1#2#3{Phys. Lett. B #1 (#3) #2}
\def\prc#1#2#3{Phys. Rev. C #1 (#3) #2}
\def\prd#1#2#3{Phys. Rev. D #1 (#3) #2}
\def\prl#1#2#3{Phys. Rev. Lett. #1 (#3) #2}
\def\phr#1#2#3{Phys. Rep. #1 (#3) #2}

\def\ptp#1#2#3{Prog. Theor. Phys. #1 (#3) #2}

\def\rmp#1#2#3{Rev. Mod. Phys. #1 (#3) #2}

\def\spjetp#1#2#3{Sov. Phys. JETP #1 (#3) #2}
\def\spu#1#2#3{Sov. Phys. Usp. #1 (#3) #2}
\def\ufn#1#2#3{Usp. Fiz. Nauk. #1 (#3) #2}
\def\yf#1#2#3{Yad. Fiz. #1 (#3) #2}
\def\zetf#1#2#3{Zh. Eksp. Teor. Fiz. #1 (#3) #2}

\newcommand{\Slash}[1]{{\ooalign{\hfil/\hfil\crcr$#1$}}}
\newcommand{\tr}{{\rm tr}}
\newcommand{\Nc}{N_{\rm c}}
\newcommand{\Nf}{N_{\rm f}}
\newcommand{\muq}{\mu}
\newcommand{\lqcd}{\Lambda_{\rm QCD}}
\newcommand{\np}{\ :\!\!}
\newcommand{\pn}{\!\!:\ }

\begin{document}
\vspace*{-5mm}
\begin{flushright}
{\scriptsize RBRC 823, BNL-90871-2009-JA, KUNS-2246}
\end{flushright}
\vspace{5mm}
\begin{frontmatter}
\title{Quarkyonic Chiral Spirals}

\author[riken]{Toru Kojo},
\author[kyoto]{Yoshimasa Hidaka},
\author[riken,bnl]{Larry McLerran}, and
\author[bnl]{Robert D. Pisarski}
\address[riken]{RIKEN/BNL Research Center, Brookhaven National
  Laboratory,\\ Upton, NY-11973, USA}
\address[kyoto]{Department of Physics, Kyoto University, Sakyo-ku, Kyoto 606-8502, Japan}
\address[bnl]{Department of Physics, Brookhaven National Laboratory, Upton,
  NY-11973, USA}

\begin{abstract}
We consider the formation of chiral density waves in Quarkyonic matter,
which is a phase where cold, dense quarks experience confining forces.
We model confinement following Gribov and Zwanziger,
taking the gluon propagator, in Coulomb gauge and momentum space, as
$\sim 1/(\vec{p}^{\; 2})^2$.
We assume that the number of colors, $\Nc$, is large,
and that the quark chemical potential, $\muq$, is much
larger than renormalization mass scale, $\lqcd$.
To leading order in $1/\Nc$ and $\lqcd/\muq$, a gauge theory
with $\Nf$ flavors of massless quarks in $3+1$ dimensions
naturally reduces to a gauge theory in $1+1$ dimensions,
with an enlarged flavor symmetry of $SU(2 \Nf)$.
Through an anomalous chiral rotation, in two dimensions
a Fermi sea of massless quarks maps
directly onto the corresponding theory in vacuum.
A chiral condensate forms locally,
and varies with the spatial position, $z$, as
$\langle \bar{\psi} \exp(2 i \muq z \gamma^0\gamma^z ) \psi \rangle$.
Following Sch\"on and Thies, we term this two dimensional pion condensate
a (Quarkyonic) chiral spiral.
Massive quarks also exhibit chiral spirals, with
the magnitude of the oscillations decreasing smoothly with increasing
mass.  The power law correlations of the 
Wess-Zumino-Novikov-Witten model in $1+1$ dimensions
then generate strong infrared effects in $3+1$ dimensions.
\end{abstract}
\end{frontmatter}


\section{Introduction}
\label{intro}
The phases of Quantum Chromodynamics (QCD) at nonzero temperature and
density are a subject of continuing interest.  While numerical simulations
on the lattice can be of use at nonzero temperature when the quark
density is small, standard Monte Carlo techniques
are not of use in cold, dense quark matter.

One expansion which is of utility is to expand in the limit of a large
number of colors \cite{thoofta,largeNreview,largeN_lattice,thooftb,2dqcd}. 
For cold, dense quark matter ---
quarks in the fundamental representation,
coupled to an $SU(\Nc)$ gauge theory ---
this gives a ``Quarkyonic'' phase 
\cite{mclerran,quarkyonic,glozman,szczepaniak}.
Keeping the quark chemical potential, $\muq$, of order one
as the number of color $\Nc \rightarrow \infty$, the free energy
for this phase is dominated by that of quarks.
Nonetheless, excitations near
the Fermi surface are confined, perhaps baryonic, whence the name.

In this paper, we consider chiral symmetry breaking in Quarkyonic matter.
We consider a phenomenological model for confinement,
taking the timelike component of the gluon propagator
to be $D^{00} \sim 1/(\vec{p}^{\; 2})^2$.
This is valid in Coulomb gauge, for a spatial momentum $\vec{p}$,
and corresponds to a potential which rises linearly in coordinate space.
Such a propagator was originally suggested by 
Gribov \cite{gribov} and Zwanziger \cite{zwanziger}.
To use such a propagator in cold, dense, quark matter, it is necessary
to assume that gluons are insensitive to screening by quarks.
For this to be true, the number of flavors, $\Nf$, must be
$\ll \Nc$, and the chemical potential must satisfy
$\muq \ll \Nc^{1/2} \lqcd$, where $\lqcd$ is the renormalization mass scale
of QCD \cite{mclerran}.  

Chiral symmetry breaking in such a model has been studied
by Glozman and Wagenbrunn \cite{glozman} and by
Guo and Szczepaniak \cite{szczepaniak}, for values of $\mu \sim \lqcd$.
We work in the extreme Quarkyonic limit,
$\muq \gg \lqcd$, so that the effects of chiral symmetry breaking
in vacuum can be ignored.  It is possible for chiral symmetry breaking
to occur at large $\muq$, since we are, by assumption, in a confined regime.
Of course there is no guarantee that our results 
apply to QCD, where $\Nc = \Nf = 3$; nevertheless,
there is certainly some range of $\Nc$, $\Nf$, and $\muq$, where it does.
If applicable to QCD, our results are of interest to intermediate
densities, where
both conventional nuclear physics and perturbative treatments fail.
Notably, this may include the astrophysics of neutron stars.

In vacuum, chiral symmetry breaking occurs through
the pairing of a left handed quark with a right handed anti-quark,
$\langle \overline{\psi}_R \psi_L\rangle \neq 0$, and
vice versa.  This condensate is, of course,
spatially uniform, so that the spontaneous breaking of chiral symmetry
does not disturb the Lorentz invariance of the vacuum.

\begin{figure}
\begin{center}
\vspace{0.5cm}
\scalebox{1.3}[1.0] {
  \includegraphics[scale=.2]{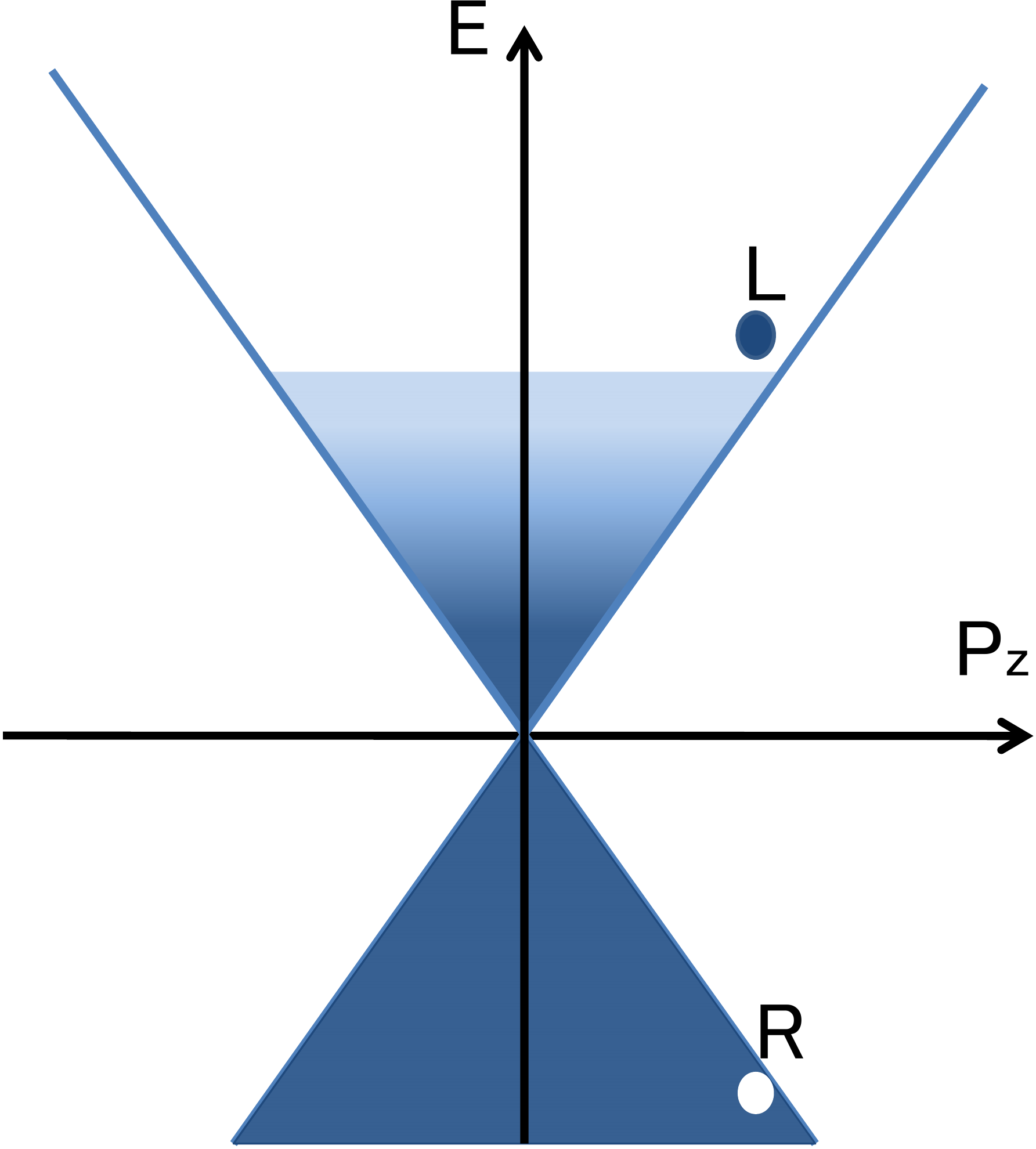} }
\end{center}
\caption{Quark anti-quark condensate in the presence of a Fermi sea.
The energy of the pair is $E_{\rm tot} \simeq 2\mu$, while its momentum is
$\vec{P}_{\rm tot}\simeq \vec{0}$.}
\label{qqbar}
\end{figure}

Now consider the effects of a Fermi sea,
where there is a net excess of quarks over anti-quarks.  The
analogy of the usual condensate is illustrated in Fig. \ref{qqbar}.
Energetically, it costs essentially 
zero energy to excite a quark right at the 
edge of the Fermi sea.  On the other hand, it costs at least
$\sim 2 \muq$ to pull an anti-quark out from deep in the Dirac sea.
(Remember that we assume that $\muq$ is very large.)
Thus the usual condensate can not be formed spontaneously, 
and anti-quarks will not enter 
into our analysis henceforth.

There are numerous features which are not captured by the illustration 
in Fig. \ref{qqbar}.  We really should draw not one, but two Fermi seas:
one for left handed quarks, and one for right handed quarks.
To avoid unnecessary duplication, instead we assume that the quark,
denoted by a filled circle, is always left handed, and that the anti-quark,
denoted by an open circle, is right handed.  The quark and anti-quark
are also assumed to have the same color, so that any condensate
is $\sim \Nc$, and survives in the limit of large $\Nc$.
If the quark has momentum $\vec{p}$, then the anti-quark,
formed by removing a quark with momentum $\vec{p}$ from the Dirac
sea, has momentum $-\vec{p}$.  
Thus the quark anti-quark pair has no net momentum, and this
condensate is spatially uniform, as in vacuum.

\begin{figure}[h]
\begin{center}
\vspace{0.5cm}
\scalebox{1.3}[1.0] {
  \includegraphics[scale=.2]{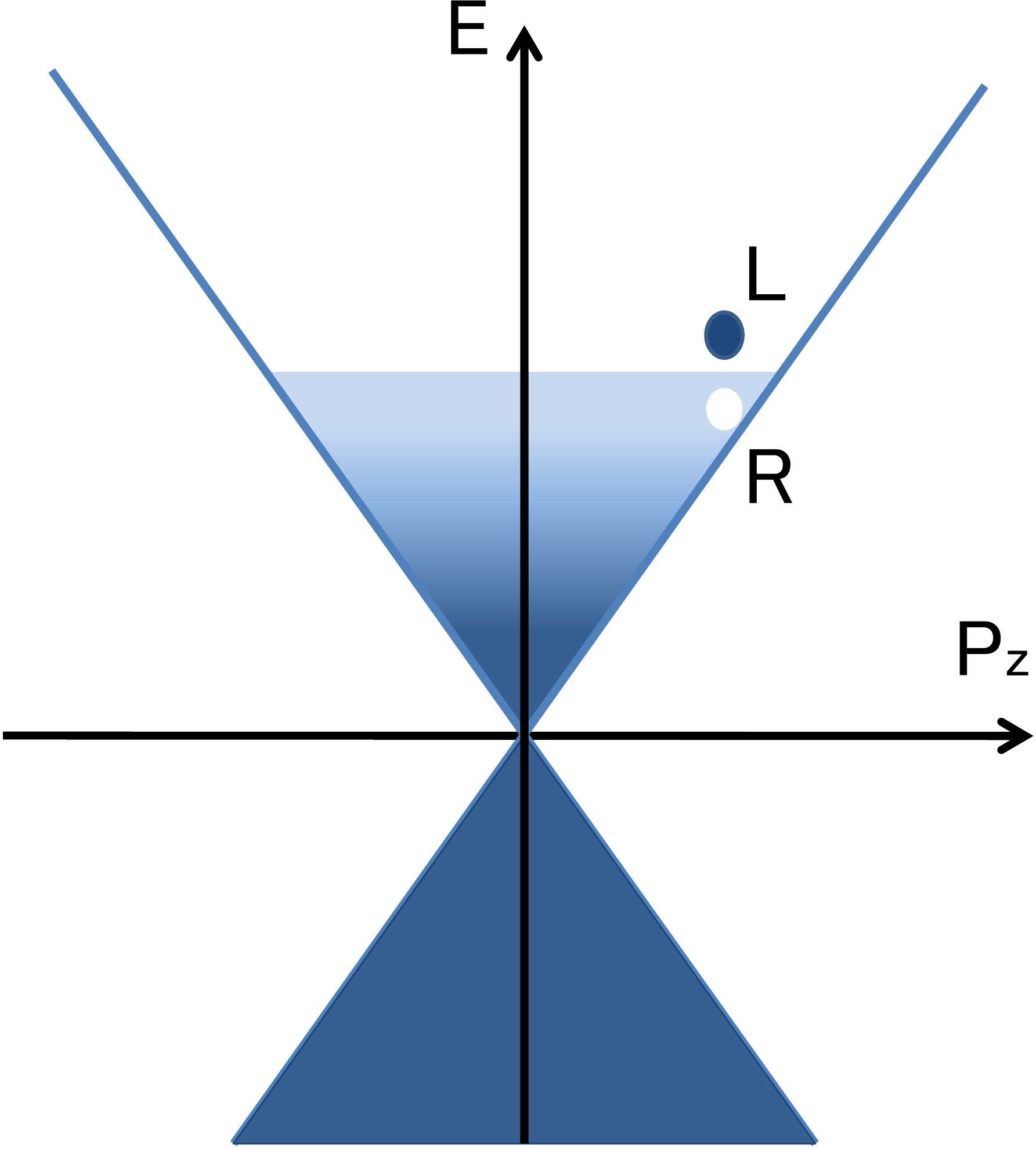} }
\end{center}
\caption{Exciton pairing between a quark and a quark hole;
$E_{\rm tot} \simeq \vec{P}_{\rm tot}\simeq \vec{0}$.}
\label{exciton}
\end{figure}

In the presence of a Fermi sea, though, it is also possible for
chiral symmetry to be broken by pairing, say, a (left-handed)
quark and a (right-handed) quark hole.
If both the quark and the quark hole are near the edge of the Fermi surface,
then it costs little energy to excite them, and the energetic penalty paid
to excite an antiquark can be avoided.
The natural analogy to the condensate in vacuum is illustrated
in Fig. \ref{exciton}, pairing a quark with momentum $\vec{p}$, and
a hole, formed by 
removing a quark with momentum $\vec{p}$ from the Fermi surface.
The momentum of the hole is then
$-\vec{p}$, so the quark-hole pair has no net momentum, and is
spatially constant.  In condensed matter physics, an excitation
as in Fig. \ref{exciton} is known as an exciton.  
Naively, we might expect that excitons are suppressed,
since the relative momentum between the particle and the hole,
$2 \vec{p}$, is large.

However, this is not the only way for quarks and their holes to break
the chiral symmetry.  Consider pairing a (left handed) quark, with 
momentum $\vec{p}$, and the hole formed by removing a (right handed)
quark with the {\it opposite} momentum, $-\vec{p}$, from the Fermi sea.
The quark hole then has the same momentum as the quark, $+\vec{p}$,
so that the resulting condensate is {\it not} uniform, and has a net
momentum $+ 2 \vec{p}$; this is, it varies 
as $\sim \exp( 2 i \muq z)$, where $\hat{z}$ is the
direction along which the pair moves, $\vec{p} = p \hat{z}$.
Such condensates do not occur in vacuum, where they 
would imply the spontaneous breaking of rotational symmetry.  
In condensed matter physics, though, such non-uniform condensates are
common, and known as density waves \cite{tsvelik}; 
this is then a chiral density wave.  Note that the
relative momentum between the quark and its hole is small, so
such a condensate may be favored.

\begin{figure}[h]
\begin{center}
\vspace{0.5cm}
\scalebox{1.3}[1.0] {
  \includegraphics[scale=.2]{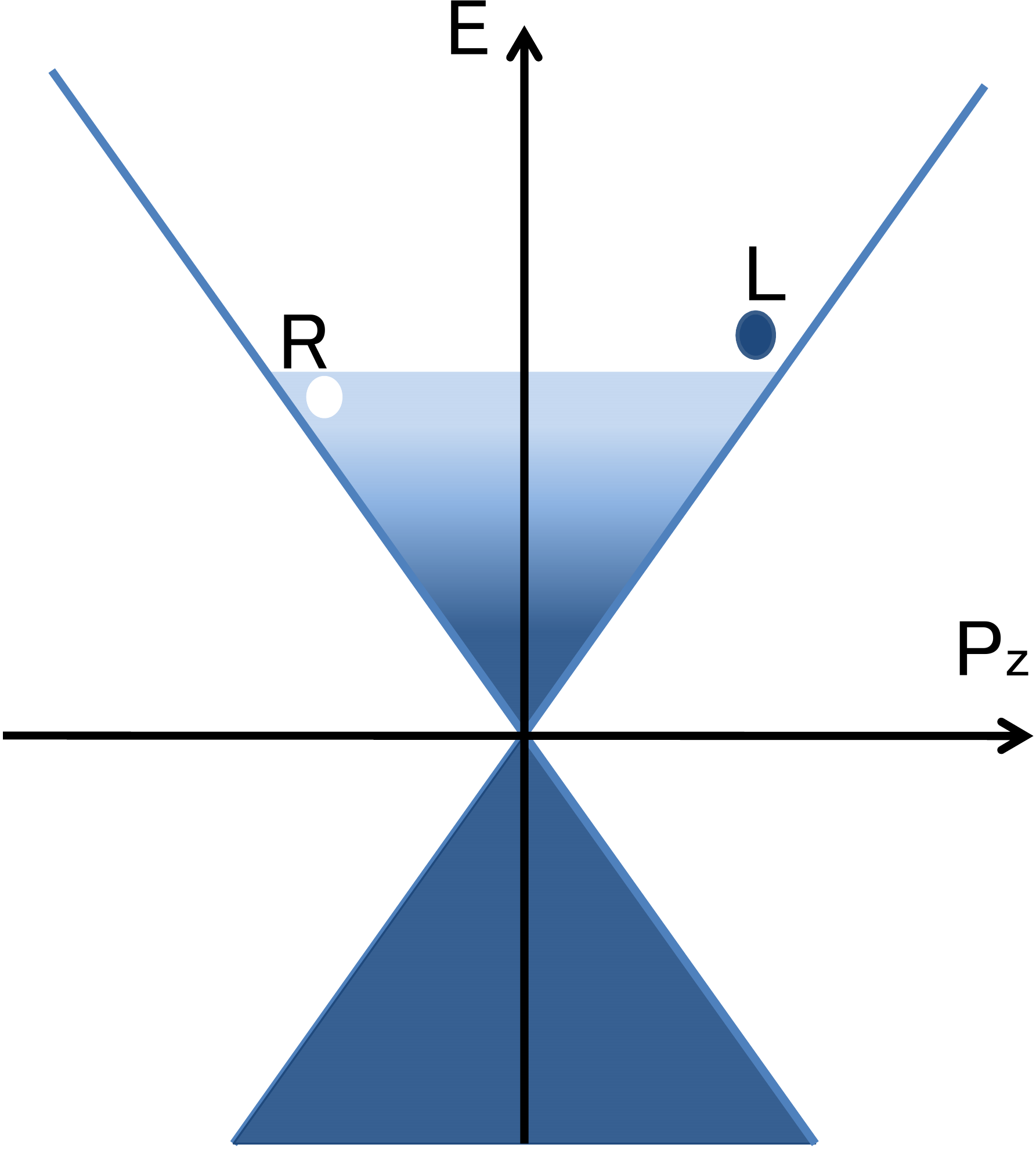} }
\end{center}
\caption{Pairing between a quark and a quark hole, which generates
a chiral density wave;
$E_{\rm tot} \simeq 0$, $\vec{P}_{\rm tot}\simeq 2 \mu \hat{z}$.}
\label{cdw}
\end{figure}

In this paper we show that in the Gribov-Zwanziger
model, that the
exciton pairing of Fig. \ref{exciton} is not generated, 
but that the chiral density wave of Fig. \ref{cdw} is.  
Again, this is familiar from systems in condensed matter: 
typically excitons are only created dynamically as resonances, 
such as by the absorption of light, and usually do not condense.
Density waves are common, especially for systems in $1+1$ dimensions
\cite{tsvelik}.  We will investigate all Dirac and flavor structures,
and show which types are preferred.

\begin{figure}[h]
\begin{center}
\vspace{0.5cm}
\scalebox{1.3}[1.0] {
  \includegraphics[scale=.2]{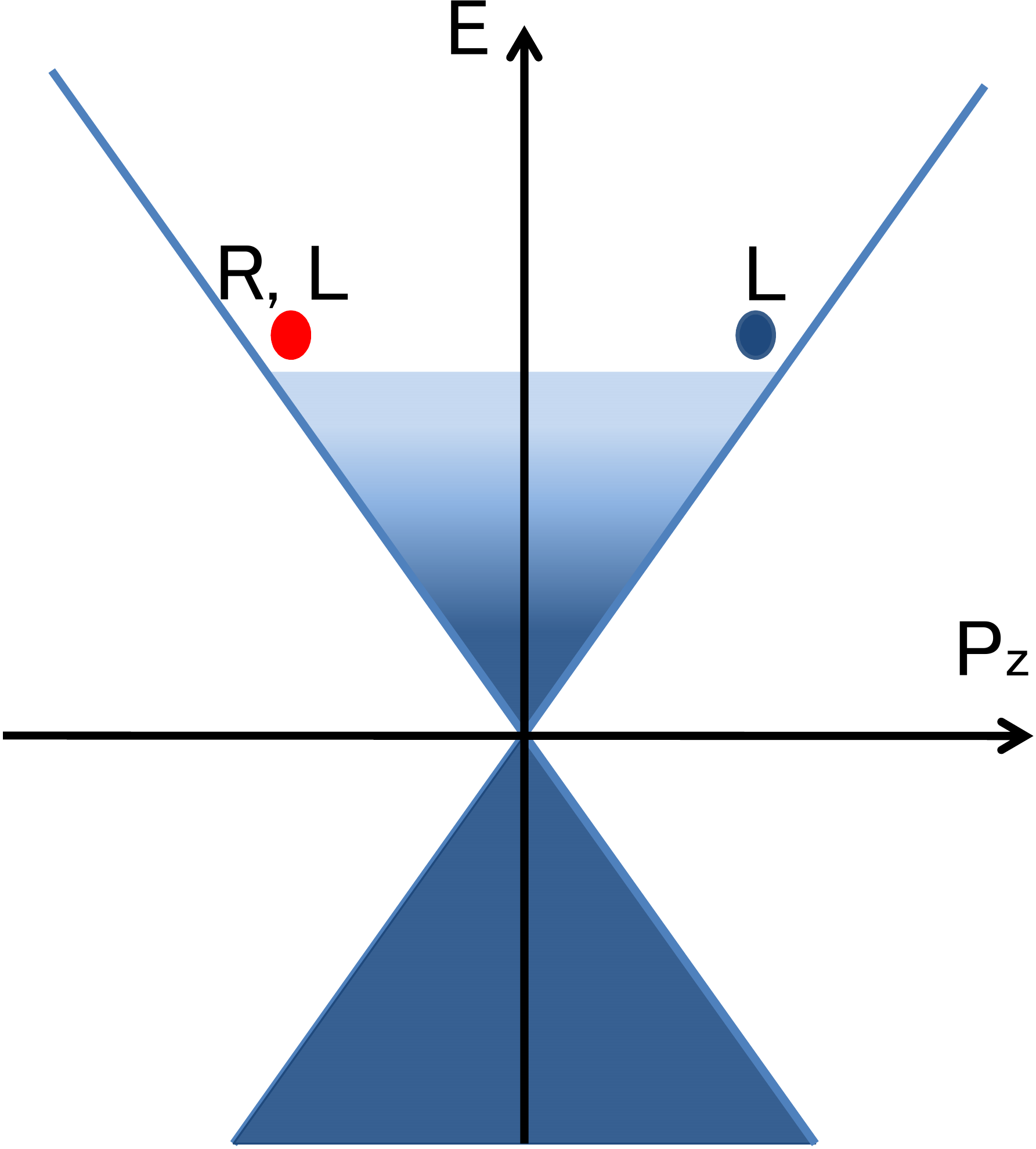} }
\end{center}
\caption{Pairing between two quarks which generates color
superconductivity;
$E_{\rm tot} \simeq \vec{P}_{\rm tot}\simeq \vec{0}$.}
\label{clr_super}
\end{figure}

For completeness, we illustrate the pairing between two quarks which
leads to color superconductivity in Fig. \ref{clr_super}.  
This is pairing between a quark at one edge of the Fermi surface,
with momentum $+\vec{p}$, and another quark at the other edge, with
momentum $-\vec{p}$.  Since pairing is between two quarks, the condensate
has no net momentum and is spatially uniform.  
For this reason, pairing can occur over the
entire Fermi surface, in a spatially symmetric state.

So far, we have not emphasized the $\Nc$ and $\Nf$ dependence
of pairing, which is not captured by the
illustrations in Figs. \ref{qqbar} - \ref{clr_super}. 
The pairing in Figs. \ref{exciton} and \ref{cdw}
is between a quark and a quark hole of the same color, so the
condensate is $\sim \Nc$. 
Further, to the extent that $\Nf \ll \Nc$, the condensate is
rather insensitive to $\Nf$.
In contrast, the diquark pairing of color superconductivity 
depends upon $\Nc$ and $\Nf$ in an essential way.
Fermi statistics greatly constrains the pairing between two quarks
(or two quark holes): it is always anti-symmetric in color,
so there are strong relations between the 
spatial wavefunction, flavor, and chirality.
For instance, in case of $\Nf=2$ and $\Nc=3$, spatially
symmetric condensates form by anti-symmetrizing in flavor; this
condensate pairs quarks of the same chirality together, and so does
not break the chiral symmetry.
On the other hand,
for $\Nf=3$ and $\Nc=3$, the preferred condensate
does break the chiral symmetry, through color-flavor locking
\cite{color_super}.

For more than three colors, the gaps for color superconductivity 
depend sensitively upon which representation one assumes the quarks 
to lie in.  If the quarks are in the fundamental representation,
then since the pairing
for color superconductivity is anti-symmetric in the
colors of the two quarks, the gap is not a color singlet, and 
is suppressed at large $\Nc$.
It is also possible, however, to generalize QCD by letting
the quarks lie in the two-index, anti-symmetric representation of color
\cite{cohen}.
This limit is rather different from that which we consider in this
paper.  There are $\sim \Nc^2$ quarks in this limit,
so that gluons are affected the quarks,
and there is no Quarkyonic phase.  This is like taking the number of
flavors, $\Nf$, to grow with $\Nc$.  In such a limit color superconductivity
is not suppressed at large $\Nc$.

It is not clear which of these two limits is most like QCD, with three
colors and three light flavors.  We suggest that it is useful to consider
all possible limits, and to
see what qualitative conclusions might be tested in QCD.

If a channel for color superconductivity exists, then Cooper pairs
will form for arbitrarily weak coupling.  Thus 
color superconductivity is always the dominant pairing mechanism at
asymptotically large chemical potential.  
The essential question is then, how large
does the chemical potential
have to be for color superconductivity to win out over other
pairing mechanisms, such as chiral density waves?

That chiral density waves \cite{tsvelik} dominate at large $\Nc$ was 
first demonstrated by Deryagin, Grigoriev, and Rubakov 
\cite{dgr,son,other_cdw,fischler,thies,bringoltz,sakai_sugimoto,pion_condensation,kaon_condensation,nickel}.  
Using a perturbative gluon propagator, $\sim 1/p^2$ in momentum space,
they find that chiral density waves form, with a condensate
\begin{equation}
\Delta_{\rm pert} \sim \mu \exp\left(-\frac{\pi}{2} 
\sqrt{\frac{\pi}{\alpha_s \Nc}} \right)
\label{perturbative_gap}
\end{equation}
in magnitude,
where $\alpha_s = g^2/4 \pi$ is the QCD fine structure constant, measured
at a scale $\mu$.
Implicitly, the computation assumes that dense quarks form a Fermi
liquid, so that pairing is from quarks (and holes) within
$\Delta_{\rm pert}$ of the edge of the Fermi sea.

In contrast, Quarkyonic matter is not a Fermi liquid because of confinement of quarks.
Low energy excitations, within
$\sim \lqcd$ of the edge of the Fermi sea, interact not through
the perturbative gluon propagator, but through the Gribov-Zwanziger form,
$\sim 1/(\vec{p}^{\; 2})^2$.  A chiral density wave forms, with the
same Dirac structure as in perturbation theory.  Because the relevant
momenta for pairing is controlled by a confining potential,
the Quarkyonic condensate for chiral density waves 
is inevitably $\sim \lqcd$ in magnitude.

Our analysis applies in a regime of intermediate $\muq$,
where the Quarkyonic gap is greater in magnitude
than the perturbative gap. 
we find that for $\Delta_{\rm pert} < \lqcd$ whenever 
$\alpha_s > 0.12$, 
using $\lqcd \sim \mu \exp(- 6 \pi/(11 \Nc \alpha_s))$.
For the coupling in QCD, this means that $\mu < 80$~GeV.
At larger values of $\mu$, $\Delta_{\rm pert} > \lqcd$.
As we discuss at the end of Sec. 2.1, this is a difficult regime to
treat, as the effects of both perturbative interactions,
and confinement, must be included.

It is known that for a Fermi liquid of dense quarks,
chiral density waves lose out to color
superconductivity except for very large values of the color,
$\Nc \geq 1000 \, \Nf$ \cite{son,other_cdw}.  This is because the
chiral density waves generated by 
perturbative interactions are very sensitive
to screening by dynamical quarks.

In sharp contrast, we expect that Quarkyonic chiral density waves 
are much less sensitive to screening by dynamical quarks.
At large $\Nc$, dynamical quarks do not affect a Quarkyonic phase
until asymptotically large values of the chemical potential,
$\mu \sim \Nc^{1/2} \lqcd$.  This power of $\Nc^{1/2}$ follows either from 
considering the Debye mass, as in Ref. \cite{mclerran},
or the free energy, as discussed in Appendix \ref{appendixB}.
As long as there is a Quarkyonic phase, and $\Delta_{\rm pert} < \lqcd$,
we expect that Quarkyonic chiral density waves dominate.

The outline of the paper is as follows.  
For quark and quark hole excitations near the Fermi surface,
where the magnitude of the transverse momentum
$|\vec{p}_\perp| \ll \mu$, the theory in $3+1$ dimensions
reduces to an effective model in $1+1$ dimensions \cite{tsvelik,son}.  
In Sec. \ref{dim_red} we show that in the Gribov-Zwanziger model,
this implies that in the gluon propagator,
we can integrate over $\vec{p}_\perp$ to obtain
\begin{equation}
\int d^2 p_\perp \; \frac{1}{(p_0^2 - p_z^2 - \vec{p}_\perp^{\; 2})^2}
\sim \frac{1}{p_0^2 - p_z^2} \; .
\label{simple_dim_red}
\end{equation}
This is the gluon propagator in $1+1$ dimensions, so our
effective theory is just QCD in $1+1$ dimensions.  
We discuss how in the Gribov-Zwanziger model,
it is necessary for $\mu \gg \lqcd$ for this reduction to hold.

In Sec. \ref{eff_lag} we consider how quantum numbers in $3+1$ dimensions
map onto those in $1+1$ dimensions.
Starting with left and right handed massless quarks
in $3+1$ dimensions, we find that the reduced model has a
doubled flavor symmetry: $\Nf$ flavors in $3+1$ dimensions becomes
an $SU(2 \Nf)$ symmetry in $1+1$ dimensions.  This extended
symmetry follows immediately from the analysis of 
Shuster and Son \cite{son}, and is very much like the doubling of
flavor symmetry which occurs for heavy quarks. 

In Sec. \ref{2D_ferm} we show how in $1+1$ dimensions,
through an anomalous redefinition of the quark fields,
a theory at nonzero chemical potential can be mapped
onto the corresponding theory in vacuum.
The net quark number, present in the theory at $\mu \neq 0$,
is generated by the axial anomaly, as shown in Appendix \ref{appendixA}.
The mapping can then be reversed: knowing results for
a gauge theory in $1+1$ dimensions in vacuum, one can read off what
happens in a Fermi sea, $\mu \neq 0$.
We show that a constant condensate in vacuum, 
$\langle \overline{\psi} \psi \rangle \neq 0$, produces
a condensate 
$\langle \overline{\psi} \exp(2 i \mu z \Gamma^5 )\psi \rangle \neq 0$, 
where $z$ is the spatial coordinate, and
$\Gamma^5$ the Dirac matrix in $1+1$ dimensions.
In the reduced model, the two dimensional $\Gamma^5 = \gamma_0 \gamma_z$,
where $\gamma_0$ and $\gamma_z$ are Dirac matrices in $3+1$ dimensions.

This type of spatially dependent condensate at $\mu \neq 0$
is familiar from soluble models in $1+1$ dimensions
\cite{fischler}, where
Sch\"on and Thies termed it a ``chiral spiral'' \cite{thies}.  A chiral
spiral was also found by Bringoltz, in his numerical analysis of
heavy quarks at nonzero density in QCD in $1+1$ dimensions
\cite{bringoltz}.  We thus term our solution a Quarkyonic Chiral Spiral
(QCS).

While we concentrate on massless quarks,
in Sec. \ref{massive_quarks} we discuss how massive quarks also
exhibit chiral spirals.  This is because
even for massive quarks, excitations about the Fermi surface are
gapless at tree level.

Effective theories about a QCS
are considered in Sec. \ref{WZW}.  
Using non-Abelian bosonization, the reduced model in $1+1$ dimensions
reduces to a Wess-Zumino-Novikov-Witten (WZNW) model 
\cite{wzw1,wzw2,wzw3}.
This model has long range correlations, which should also produce
long range correlations in $3+1$ dimensions.

We conclude in Sec. \ref{conclusions} about whether QCS's may be be relevant
for cold, dense QCD.  We note that QCS's are closely related to
pion condensation \cite{pion_condensation}.
They are not identical, because a pion condensate is a chiral spiral in
the four dimensional $\gamma_5$, while for a QCS, it is in the two
dimensional $\Gamma^5$.  We also note that if pionic chiral spirals occur,
then probably so do kaonic chiral spirals.  As a spatially varying condensate,
a kaonic chiral spiral differs for the spatially constant condensate of kaon
condensation \cite{kaon_condensation}.
Chiral density waves 
also arise in the Sakai-Sugimoto model \cite{sakai_sugimoto}.

The physics of QCS's
should be especially rich, however, since it
includes the spontaneous breaking
of translational and rotational 
symmetries, and a plethora of light modes.  Such phenomenon
should have direct implications for observations of 
neutron/Quarkyonic stars.

\section{Dimensional Reduction: Self-Consistent Equations}
\label{dim_red}

\subsection{Reduction of the Schwinger-Dyson Equation}
\label{red_sch_dy}

We start by considering the Schwinger-Dyson equation
for the quark self-energy.
In a Quarkyonic phase at large $\Nc$, the gluons are unaffected
by the quarks, so that corrections to the gluon self-energy,
and vertices, can be neglected.  
For the quark self-energy we take the sum of rainbow diagrams,
\begin{equation}
\Slash{\Sigma}(p) 
= - \int \frac{d^4 k}{(2\pi)^4} \; D^{AB}_{\mu \nu}(p-k)\;
 (\gamma_\mu t_A) \; S(k) \; (\gamma_\nu t_B) \; ,
\label{SD1}
\end{equation}
where $\Sigma(p)$ is the quark self-energy.  At a nonzero chemical
potential $\mu$, the dressed quark propagator, $S(k)$, is 
\begin{equation}
S(k) 
= \frac{1}{[k_4+i\mu+\Sigma_4(k)]\gamma_4 
+ [k_j + \Sigma_j(k)]\gamma_j } \; .
\label{SD2}
\end{equation}
where $\Sigma$ is determined self-consistently through the
integral equation, Eq. (\ref{SD1}).  We work in Euclidean
spacetime, $k = (k_4, \vec{k})$.  In the limit that $\mu \gg \lqcd$,
we neglect chiral symmetry effects as in vacuum, as illustrated in
Fig. \ref{qqbar}; such effects have been considered at $\mu \neq 0$
by Refs. \cite{glozman,szczepaniak}.  Consequently, we neglect terms
$\sim 1$ in the quark self-energy and propagator.

For the gluon propagator, $D_{\mu \nu}(k)$, we take the
Gribov-Zwanziger form,
\begin{equation}
D_{44}^{A B}(k) 
= - \frac{8\pi}{C_F} \times \frac{\sigma }{(\vec{k}^2)^2} \; \delta^{A B}
\; \; \; ; \;\;\;  
D^{4 i} = D^{ij} = 0 \,.
\label{gluon_propagator}
\end{equation}
At the outset, we stress that we are dealing with a {\it model}
of confinement.  While the gluon propagator,
and vertices, are unaffected by quark loops, there is no fundamental
justification in taking the vertices to be the same as the bare ones,
nor in taking the gluon propagator of Eq. (\ref{gluon_propagator}).
The gluon propagator involves a parameter, $\sigma$, which is the
string tension, and has dimensions of mass squared, $\sigma \sim \lqcd^2$.
Numerical factors $8\pi$ and $C_F=(\Nc^2-1)/2\Nc$
are multiplied to
reproduce a correct linear potential for the color singlet channel.

We stress that the propagator in Eq. (\ref{gluon_propagator}) is valid 
only for {\it small} momenta, for $k \lesssim \lqcd$.  For larger
momenta, one should use the usual gluon propagator of perturbation
theory, $\sim 1/k^2$.  
For excitations near the edge of the Fermi sea, though, we can
neglect the perturbative part of the propagator.
This differs, for example, from the computation of the free energy
\cite{mclerran}.  That is dominated by momenta transfers within the entire
Fermi sea, $\sim \mu$, for which the perturbative gluon (and
quark) propagators should be used.  There are contributions to the free
energy from momenta $\sim \lqcd$, but this are small, powers
of $\sim (\lqcd/\mu)^2$ times the perturbative terms 
\cite{mclerran,quarkyonic}.  

After summing over the color indices, Eq. (\ref{SD1}) becomes
\begin{equation}
\Slash{\Sigma}(p) 
=  \int \frac{d^4 k}{(2\pi)^4} \; 
\frac{8\pi \sigma}{((\vec{p}-\vec{k})^2)^2} \;
\gamma_4 \; S(k) \; \gamma_4 \; ;
\label{SD3}
\end{equation}
normalizing the generators as $\tr(t^A t^B) = \delta^{A B}/2$.
Since the right hand side of this equation is independent of $p_4$,
so is the quark self-energy, $\Slash{\Sigma}(p)$.

The Schwinger-Dyson equation simplifies considerably if we consider
only excitations near the edge of a Fermi sea.  
For a free massless quark with momentum 
$p^\mu=(p_4,p_z,p_y,p_x)=(p_4,p_z,\vec{p}_\perp)$,
in a Fermi sea its mass shell is given by 
\begin{equation}
i p_4 = \sqrt{p_z^2 + \vec{p}_\perp^{\; 2} } - \mu \,.
\label{mass_shell1}
\end{equation}
Assuming that quark is along the $\hat{z}$ direction, so that
$|p_z| = \mu + \delta p_z$, 
\begin{equation}
i p_4 \approx \delta p_z + \frac{\vec{p}_\perp^{\; 2} }{2 \mu} + \ldots \,.
\label{mass_shell2}
\end{equation}
Thus, as is well known \cite{tsvelik}, in a Fermi sea
the dispersion relation linearizes in $\delta p_z$, allowing
us to neglect the effects of the transverse momenta for the quarks.

Since this is the mass shell for a free quark, it neglects the effects
of the quark self-energy, $\Sigma$.  
We expect, though, that for quarks and quark holes
near the Fermi surface, including the quark self-energy
does not affect the 
suppression of fluctuations in $\vec{p}_\perp$.
For the Gribov-Zwanziger potential,
the natural scale for the transverse momenta is $|\vec{p}_\perp| \sim \lqcd$.
Thus we can neglect the quark transverse momenta 
in the extreme Quarkyonic limit, where
$\mu \gg \lqcd$.  What happens when $\mu \sim \lqcd$ is a difficult problem 
which we do not address here.

In the Schwinger-Dyson equation, the dominant contribution from such
an infrared singular gluon propagator is when the gluon momentum is small.
This constrains the internal and external momenta of the quarks
to be near one another, $k \approx p$.  Neglecting the transverse momenta
of the quark, $\vec{k}_\perp$, we need only consider the two components
of the momenta along the light cone, $k_z$ and $k_4$.  
The Schwinger-Dyson equation thus reduces to
\begin{equation}
\Slash{\Sigma} (p) \simeq 
\int \frac{d^4 k}{(2\pi)^4}\;
\gamma_4 \; S(k_4,k_z,\vec{0}_\perp) \; \gamma_4
\frac{8 \pi \sigma}{\left((\vec{k}_\perp-\vec{p}_\perp)^2 + (k_z-p_z)^2\right)^2}
\; .
\label{SD3b}
\end{equation}
Since the transverse momentum $\vec{k}_\perp$ enters only through
the gluon propagator, we can now integrate it out,
\begin{equation}
\int \frac{d^2 k_\perp}{(2\pi)^2} 
\frac{8\pi \sigma}
 {\left((\vec{k}_\perp-\vec{p}_\perp)^2 + (k_z-p_z)^2\right)^2} 
= \frac{\Nc g^2_{\rm 2D} }{2} \; \frac{1}{(k_z - p_z)^2} \;\;\; ; \;\;\;
\Nc g^2_{\rm 2D} = 4\sigma \; ,
\label{SD4}
\end{equation}
where we define  a two dimensional gauge coupling constant
$g_{{\rm 2D}}$, and neglect $\vec{p}_\perp$.

The reduced Schwinger-Dyson equation then becomes
\begin{equation}
\Slash{\Sigma} (p_4,p_z, \vec{0}_{\perp}) \simeq 
\frac{\Nc g^{2}_{{\rm 2D}} }{2} \; \int \frac{dk_4 \, dk_z}{(2\pi)^2}\;
\gamma_4 \; S(k_4,k_z, \vec{0}_{\perp}) \; \gamma_4 \; 
\frac{1}{\left(k_z-p_z\right)^2} \,.
\label{SD5}
\end{equation}

This integral equation corresponds to QCD in $1+1$ dimensions.
In axial gauge, $A_z = 0$, the two dimensional action for gluons
reduces to a free term,
\begin{equation}
\frac{1}{2 }
\tr \; G_{\mu \nu}^2
= 
\tr\;({\partial_z} A_4)^2 \; .
\end{equation}
In the limit of large $\Nc$, the model in $1+1$ dimensions
obviously gives the integral equation of Eq. (\ref{SD5}).

Of course it is necessary to take some care in the reduction of the
Dirac matrices from $3+1$ to $1+1$ dimensions.  We address this
in Sec. \ref{eff_lag}.

A more realistic model for the gluon propagator than
Eq. (\ref{gluon_propagator}) is to take 
a sum of the Gribov-Zwanziger term,
$\sim 1/(\vec{k}^2)^2$, plus a perturbative piece, $\sim 1/k^2$.  
The confining term is valid for momenta $\lesssim \lqcd$; the perturbative term,
for momenta $> \lqcd$.  
The analysis goes through as above.  
The dependence of the quark propagators on the transverse
momenta can be neglected, so one is left with an integral of the
gluon propagator with respect to $\vec{k}_\perp$.
Integration over the Gribov-Zwanziger propagator gives 
$\sim \sigma/(k_z-p_z)^2$, while the integral
over the perturbative piece generates a logarithm \cite{son,other_cdw},
$\sim C_F g^2 \log(\mu^2/(k_z-p_z)^2)$.

With only a perturbative propagator, $\sigma = 0$, the analysis is as follows
\cite{dgr,son,other_cdw}:
Assuming that the $\Sigma_m$ is a constant gap,
$\Delta_{\rm pert}$, 
the dimensional reduction can apply to the region
of quark momenta,
$\Delta_{\rm pert}^2/\mu \ll |\delta k_z| \ll \Delta_{\rm pert}$
($\delta k_z$ is measured from the Fermi momentum).
Integrating over $k_4$ of a quark propagator
yields a factor of the inverse energy,
$1/|\delta k_z^2 + \Delta_{\rm pert}^2|^{1/2}$.
The form of self-consistent equation has a similar structure
as the BCS gap equation for the constant gap.
A dominant contribution to the integral comes from
the inverse energy part, which is sensitive to the gap
(an absence of gap yields infrared divergence in the integrand.
This is nothing but the instability of the Fermi surface.)
Integration over $k_z$ at soft momentum region gives 
a logarithm term, $\sim\log(\mu/\Delta_\text{pert})$.
Combining it with the logarithm of the gluon propagator,
one find the squared logarithmic form of the gap equation, 
$\Delta_{\rm pert} \sim g^2 \Nc \log^2(\mu/\Delta_{\rm pert}) 
\Delta_{\rm pert}$, with
the self-consistent solution of Eq. (\ref{perturbative_gap}).

When we simply add the confining interaction
to the self-consistent equation, its nature strongly changes.
The self-energy can be no longer identified as the gap from 
the symmetry breaking, because the single-excitation-energy
gap due to the confinement potential is large or divergent.
The integral over $k_z$ in the gap equation is 
$\int dk_z 1/|\delta k_z^2 + \Sigma_m^2|^{1/2}
\times \sigma/(\delta k_z - \delta p_z)^2
 \sim \sigma/\Lambda_\text{IR} \times f(\delta p_z)$, 
where $f(\delta p_z)$ is a some
regular function and $\Lambda_\text{IR}$ is an infrared cutoff.
The infrared cutoff $\Lambda_\text{IR}$ for the gap equation for the single 
particle spectrum should be identified with the Debye mass, 
so that the gap itself is of order $\sqrt{\Nc} \lqcd^2/(\sqrt{g^2 \Nc} \mu)$.  
To see this imagine we try to ionize a hadronic state into its 
constituent single particle quark excitations.  This can only 
happen when the separation between the constituents is of 
order the Debye screening length, and costs an energy 
$\sigma R_{\rm Debye} = 
\sqrt{\Nc} \lqcd^2/(\sqrt{g^2 \Nc}\mu)$.  This provides a gauge invariant 
definition for the single particle excitation gap.  It also shows 
that the gap is large compared to the confinement scale 
in the Quarkyonic Phase. There is no weakly
coupled solution for the gap equation when in this phase since,
if one goes back to the derivation of the perturbative 
contribution, the derivation of the logarithmic terms is no 
longer valid.  The dominant contribution for the perturbative 
piece in  arises for momenta greater than the gap itself.

It is important to understand that the analysis here concerns 
only the single particle excitation spectrum.
The issue of chiral condensation is not directly related.
We have provided a self-consistent derivation of the chiral 
condensate in other parts of this paper, and  the condensate 
arises from non-perturbative effects.  In a self-consistent 
analysis of course one cannot rule out the possibility that 
there is another solution arising from a different kinematic region.

As discussed in the Introduction, we consider only a 
region of intermediate
$\mu$, where the $\Delta_{\rm pert} < \lqcd$.
In this region, it is safe to include
only the confining propagator, and neglect the perturbative term.  
When $\Delta_{\rm pert} > \lqcd$, we enter a more complicated regime,
as the large difference in momentum scales required by the perturbative
dimensional reduction, 
between $\Delta_{\rm pert}^2/\mu$ and $\Delta_{\rm pert}$,
is manifestly affected by confinement. 
Confinement can probably be neglected when $\lqcd < \Delta_{\rm pert}^2/\mu$,
but this only occurs when $\alpha_s \sim 0.03$, which is an
astronomically large value of $\mu$, $\sim 10^{11}$GeV.

\subsection{Reduction of the Bethe-Salpeter Equation}
\label{red_bethe_sal}


In this subsection we show how the reduction of the Schwinger-Dyson
equation for the quark propagator applies as well to the Bethe-Salpeter
equation for mesonic wave functions.  Henceforth we work in
Minkowski spacetime, $k=(k_0,\vec{k})$.
We consider the homogeneous equation, which in large $\Nc$ 
is of the ladder type,
\begin{equation}
\widetilde{\Psi}(P;q)_{\alpha \beta}^{ab}
= -\int \frac{d^4k}{(2\pi)^4} \;
\big[ S(k-P)(\gamma^\mu t_A) \; \widetilde{\Psi}(P;k) \; (\gamma^{\nu} t_B)
S(k+P) \big]_{\alpha \beta}^{ab}
D^{AB}_{\mu \nu}(k-q) ;
\end{equation}
$a$ and $b$ denote color indices, $\alpha$ and $\beta$ spinor indices.
$\widetilde{\Psi}(P;q)$ is a bound state wave function,
where $2P$ is the total momentum, $q$ the relative momentum.
The dressed quark propagator, $S(q)$,
is the self-consistent solution to the Schwinger-Dyson equation,
Eq. (\ref{SD5}).  The Bethe-Salpeter equation includes both
singlet and adjoint channels, 
$\tilde{\Psi}^{ab} = \delta^{ab} \Psi + t_A^{ab} \Psi_A$;
for the singlet channel, this reduces to 
($D_{00}=-D_{44}$)
\begin{equation}
\Psi(P;q)_{\alpha \beta} =
- \int \frac{d^4 k}{(2\pi)^4} 
\big[ S(k-P) \; 
\gamma^{0} \Psi(P;k) \gamma^{0} \; S(k+P) \big]_{\alpha \beta}
\frac{8\pi \sigma}{((\vec{k}-\vec{q}\,)^2)^2} \; .
\end{equation}

As in the previous subsection, we concentrate on very soft
gluons, $\vec{k} \approx \vec{q}$.  We assume that the
both quarks, with momenta $\vec{k}-\vec{P}$ and $\vec{k}+\vec{P}$ are close to
the Fermi momentum, $|\vec{p}_F| = \mu$.  
We assume that the quark and quark hole
pair have a total momentum along the $z$-direction, $2\vec{P} = 2P \hat{z}$,
and that the transverse momentum of the wavefunction can be neglected.
We consider pairing energies that are both of the exciton type,
$P_z \simeq 0$ and $|k_z|\simeq p_F$, 
and for a chiral density wave, $|P_z| \simeq p_F$ and $k_z \simeq 0$.

Neglecting the transverse momenta of the quark and quark hole,
as we did for the quark mass shell in Eq. (\ref{mass_shell2}), 
we then introduce a wavefuntion, $\varphi$, which is
obtained by averaging 
$\Psi(P;q)$ over the relative, transverse momenta, $\vec{q}_\perp$:
\begin{equation}
\varphi(P;q_0,q_z) 
= \int \frac{d^2 \vec{q}_\perp }{(2\pi)^2} 
\; \Psi(P;q_0,q_z,\vec{q}_\perp) \; .
\end{equation}
The Bethe-Salpeter equation satisfied by this averaged
wavefunction is
\begin{equation}                                                              
\begin{split}
\varphi(P;q_0,q_z)_{\alpha \beta}                                            
=&                                                                          
- \int \frac{d^2 \vec{q}_\perp }{(2\pi)^2} 
\; S(q-P)_{\alpha \gamma}S(q+P)_{\delta \beta}  \\
&\qquad \times                                                         
\int \frac{d^4k}{(2\pi)^4}                                
\; \big[ \gamma^{0} \Psi(P;k) \gamma^{0} \big]_{\gamma \delta}              
\; \frac{8\pi \sigma}{((\vec{k}-\vec{q}\,)^2)^2}.                             
\end{split}
\end{equation}

We take the momenta of the quarks to be near the Fermi surface,
$\vec{q} \pm \vec{P} = ( p_F \pm \delta p_z, \pm \delta \vec{p}_\perp)$.
Thus we introduce two dimensional quark propagators,
\begin{equation}
\bar{S}(P\pm q) \equiv S(P_0\pm q_0,P_z \pm q_z, \vec{0}_\perp) \,.
\end{equation}
The Bethe-Salpeter equation becomes
\begin{equation}
\begin{split}
\varphi(P;q_0,q_z)_{\alpha \beta} 
\simeq& 
- \bar{S}(q-P)_{\alpha \gamma} \bar{S}(q+P)_{\delta \beta}  \\
& \times
\int \frac{d^4 k}{(2\pi)^4}
\; \big[ \gamma^{0} \Psi(P;k) \gamma^{0} \big]_{\gamma \delta}
\; \int \frac{d^2 \vec{q}_\perp}{(2\pi)^2}
\; \frac{8\pi \sigma}{((\vec{k}-\vec{q}\,)^2)^2} \; .
\end{split}
\end{equation}
The wave function on the right hand side depends only upon $P$ and
$k$, but not upon $q$, and so we can integrate $\Psi(P;k)$ with
respect to $\vec{k}_\perp$.  That implies that as for the quark
self-energy, we can integrate over $\vec{q}_\perp$, leaving
\begin{equation}
\begin{split}
\varphi(P;q_0,q_z)_{\alpha \beta} 
&\simeq
- \; \frac{\Nc g^2_{{\rm 2D}}}{2} \;
\bar{S}(q-P)_{\alpha \gamma} \bar{S}(q+P)_{\delta \beta}  \\
& \quad\times\int \frac{dk_0 dk_z}{(2\pi)^2}\; 
\big[ \gamma^{0} \varphi(P;k) \gamma^{0} \big]_{\gamma \delta}
\; \frac{1}{(k_z-q_z)^2} \; .
\end{split}
\end{equation}
This is the same form as the Bethe-Salpeter equation for
QCD in two dimensions \cite{thooftb}, in $A_z = 0$ gauge.

\section{The Effective Lagrangian in $1+1$ Dimensions}
\label{eff_lag}

In the previous Section we demonstrated how dimensional reduction,
from $3+1$ to $1+1$ dimensions, occurs for 
both the quark self-energy and the Bethe-Salpeter wave function.
In this section we show how this arises at the level of effective
Lagrangians.  Our discussion elaborates upon that
by Shuster and Son \cite{son}.

Start with the free quark Lagrangian in $3+1$ dimensions.  Since
we concentrate on quarks near the Fermi surface, we assume that the
spatial momentum is along the $\hat{z}$ direction, 
\begin{equation}
{\cal L}_{\rm kin}^{\rm lightcone} =i(\bar{\psi} \gamma^0 \partial_0 \psi 
+ \bar{\psi} \gamma^z \partial_z \psi) \,.
\label{eq:LagrangianLightcone}
\end{equation}
For the Dirac matrices, we take
\begin{equation}
\gamma^0 = \left[
\begin{array}{ccc}
\ 0\ &\ {\bf 1}\  \\
{\bf 1} & 0
\end{array}
\right],\ \ 
\gamma^j = \left[
\begin{array}{ccc}
0 & -\sigma^j  \\
\sigma^j & 0
\end{array}
\right],\ \ 
\gamma^5 = \left[
\begin{array}{ccc}
\ {\bf 1}\ &\ 0\  \\
0 & -{\bf 1}
\end{array}
\right] \; ; \ 
\end{equation}
$\sigma^j$ are the Pauli matrices, and $\bf 1$ the unit matrix
in two dimensions. 
We also define spin matrices as 
\begin{equation}
\Sigma^{i}= \gamma^5 \, \gamma^0 \gamma^i =
\frac{i\epsilon^{ijk}}{4} [\gamma^{j},\gamma^{k}] =
\left[
\begin{array}{ccc}
\ \sigma^{i}\ &\ 0\  \\
0 & \sigma^{i}
\end{array}
\right] 
\, ;
\end{equation} 
$\epsilon^{ijk}$ is the totally antisymmetric tensor, $\epsilon^{123}=1$.
If transverse momenta can be neglected, 
the reduced Lagrangian in $1+1$ dimensions
has an extended symmetry,
which is related to the spin that quarks carry in $3+1$ dimensions.

We introduce projectors for the quark fields,
\begin{equation}
\psi_{R,L} = \frac{1 \pm \gamma_5}{2}\; \psi \; \; ; \;\;
\psi_{R\pm} = \frac{1 \pm \gamma^0 \gamma^z}{2}\psi_R \;\; ; \;\;
\psi_{L\pm} = \frac{1 \pm \gamma^0 \gamma^z}{2}\psi_L \; .
\label{projectors}
\end{equation}
$\psi_R$ and $\psi_L$ are right and left handed
fields, eigenstates of chirality. 
We also introduce projectors for 
spin \cite{spin} along the $\hat{z}$ direction, $(1 \pm \Sigma^z)/2$;
for eigenstates of chirality, this equals
$(1 \pm \gamma^0 \gamma^z)/2$.

The usual chiral basis is to take
\begin{equation}
\psi^T = 
\left[ \psi_{R+} \; , \;  \psi_{R-} \; , \; \psi_{L-} \; , \;  \psi_{L+}
\right] \; ,
\end{equation}
but for this problem, this is rather inconvenient.  To see this,
we write the free Lagrangian in terms of $\psi_{R\pm}$ and
$\psi_{L\pm}$, 
\begin{equation}
\begin{split}
{\cal L}_{\rm kin}^{\rm lightcone}
=& i[\psi_{R+}^\dag (\partial_0 + \partial_z) \psi_{R+}
+ \psi_{R-}^\dag (\partial_0 - \partial_z) \psi_{R-}\\
&+ \psi_{L+}^\dag (\partial_0 + \partial_z) \psi_{L+}
+ \psi_{L-}^\dag (\partial_0 - \partial_z) \psi_{L-}
] \,.
\end{split}
\end{equation}
The fields $\psi_{R+}$ and $\psi_{L-}$ have spin up along
the $+ \hat{z}$ direction, while $\psi_{L+}$ and $\psi_{R-}$ have spin 
along the $- \hat{z}$ direction.
In terms of two dimensions, for positive energy 
$\psi_{R+}$ and $\psi_{L+}$ are right moving fields,
while $\psi_{L-}$ and $\psi_{R-}$ are left moving fields.

Because there is no spin in two dimensions, we only need two component
spinors.  Thus from one four component spinor
in $3+1$ dimensions we obtain two types of two component 
spinors in $1+1$ dimensions, 
\begin{equation}
\varphi_\uparrow = 
\left[  
\begin{array}{cc}
\psi_{R+}\\  
\psi_{L-} 
\end{array}
\right] 
\;\;\; ; \;\;\;
\varphi_\downarrow = 
\left[ 
\begin{array}{cc}
\psi_{L+} \\
\psi_{R-} 
\end{array}
\right] \; .
\end{equation}
The spinors $\varphi_\uparrow$ and $\varphi_\downarrow$ 
are eigenstate of spin along the $\hat{z}$ direction,
and act as two ``flavors'' in $1+1$ dimensions.  
This is valid to the extent
that we can neglect the transverse momenta, which
couple to the flavor breaking matrix, $\vec{\gamma}_\perp$.

We can then combine $\varphi_\uparrow$
and $\varphi_\downarrow$ into one four component spinor, $\Phi$, that is
related to the original quark field, $\psi$, by a unitary transform, $U$:
\begin{equation}
\Phi = 
\left[
 \begin{array}{cc}
 \varphi_\uparrow \\
 \varphi_\downarrow
 \end{array}
\right] 
= U \; \psi \;\;\; ; \;\;\;
U = \left[
 \begin{array}{cccc}
\ 1\ &\ 0\ &\ 0\ &\ 0\ \\
 0 & 0 & 1 & 0 \\
 0 & 0 & 0 & 1 \\
 0 & 1 & 0 & 0
 \end{array}
\right] 
\quad; \quad
U^{\dagger} U = 1 \,.
\end{equation}

For the Dirac matrices in two dimensions, $\Gamma^\mu$, $\mu = 0,z$,
we make the obvious choice,
\begin{equation}
\Gamma^0 = \sigma^1 \;\;\; ; \;\;\;
\Gamma^z = - i \sigma^2 \;\;\; ; \;\;\;
\Gamma^5 = \sigma^3 \; .
\label{twodim_pauli}
\end{equation}
In two dimensions, 
\begin{equation}
\Gamma^0\, \Gamma^z = \Gamma^5 \; .
\label{twodim_identity}
\end{equation}
While this identity is trivial mathematically, it plays an
important role in the next section.  
The conjugate is defined as
$\overline{\Phi} =
(\bar{\varphi}_\uparrow,\bar{\varphi}_\downarrow) 
= (\varphi_\uparrow^\dag, \varphi_\downarrow^\dag) 
\Gamma^0$.

It is straightforward to rewrite quark bilinears in $3+1$ dimensions
in terms of spinors in $1+1$ dimensions.  The operators appearing in the
action, 
\begin{align}
\hspace{-0.75cm}
 \bar{\psi}\gamma^0 \psi 
\; =& \; \psi^\dag_{R+}\psi_{R+} 
 + \psi^\dag_{R-}\psi_{R-}
 + (R \leftrightarrow L)
\; = \; \Phi^\dag \, \Phi  
\; = \; \overline{\Phi}\; \Gamma^0 \; \Phi \; ; \nonumber \\
\bar{\psi}\gamma^z \psi 
\; =& \; \psi^\dag_{R+}\psi_{R+} 
 - \psi^\dag_{R-}\psi_{R-}
 + (R \leftrightarrow L)
\; = \; \Phi^\dag \, \Gamma^5 \, \Phi 
\; = \; \overline{\Phi}\; \Gamma^z \; \Phi \; ,
\end{align}
directly map from four to two dimensions.
Because of these identities, we see that in the presence of
gauge fields, and a nonzero chemical potential, that a gauge
theory in $3+1$ dimensions maps into one in $1+1$ dimensions,
\begin{equation}
{\cal L}^{{\rm 2d}}_{{\rm eff}} 
= \overline{\Phi} \big[
i \, \Gamma^\mu  \left(\partial_\mu + i \, g_{\rm 2d}\, A_\mu\right) 
+ \mu \, \Gamma^0 \big] \Phi
- \frac{1}{2} \; \tr \; G_{\mu\nu}^2 \,.
\end{equation}
Fermions in two dimensions only require two components.  Thus a single,
four component spinor in four dimensions
becomes two types of two components spinors in two dimensions:
$\psi$, or equivalently $\Phi$, become $\varphi_i$, 
where $i= \ \uparrow, \ \downarrow$
is the flavor index in two dimensions, generated dynamically
by dimensional reduction.

We note that the mass term reduces similarly,
\begin{equation}
\hspace{-0.75cm}
\bar{\psi} \psi 
= \psi^\dag_{R+}\psi_{L-} 
 + \psi^\dag_{L-}\psi_{R+}
 + (R \leftrightarrow L)
= \overline{\Phi} \, \Phi \,.
\label{mass_transf}
\end{equation}
One can write the complete dictionary to go from quark bilinears
in $3+1$ dimensions to those in $1+1$ dimensions.
We introduce matrices for the two dimensional flavor, 
which act in the space of $\varphi_\uparrow$ and $\varphi_\downarrow$:
\begin{equation}
{\bf \tau}_f 
= ({\bf \tau}_1, {\bf \tau}_2, {\bf \tau}_3)
=  U\,( \gamma^0 \Sigma^{1}, \gamma^0 \Sigma^{2}, \Sigma^{3} )\,U^{\dag}
= \bigg(
\left[
\begin{array}{ccc}
\ 0\ &\ {\bf 1}\  \\
{\bf 1} & 0
\end{array}
\right]
,
\left[
\begin{array}{ccc}
0 & -i {\bf 1}  \\
i{\bf 1} & 0
\end{array}
\right]
,
\left[
\begin{array}{ccc}
\ {\bf 1}\ &\ 0\ \\
0 &  -{\bf 1}
\end{array}
\right]
\bigg).
\end{equation}

For operators which are diagonal in flavor,
\begin{equation}
\overline{\Phi}
\left[
\begin{array}{ccc}
\Gamma^a & 0\\
0 & \Gamma^a
\end{array}
\right]
\Phi 
  =   \bar{\psi} \; \gamma^0 \; U^\dag
\left[
\begin{array}{ccc}
\Gamma^0 \, \Gamma^a & 0  \\
0 & \Gamma^0 \, \Gamma^a
\end{array}
\right]
U \; \psi \; ;
\end{equation}
where $\Gamma^a = {\bf 1}$, $\Gamma^5$, $\Gamma^0$, or $\Gamma^z$.

For operators which are not diagonal in flavor, we note that
the flavors matrices ${\bf \tau}_f$ do 
not mix right and left moving components, so
$\Gamma^a$ and ${\bf \tau}_f$ commute with each other.
Thus operators which are not flavor singlets transform as
\begin{equation}
\overline{\Phi}
\left[
\begin{array}{ccc}
\Gamma^a & 0\\
0 & \Gamma^a
\end{array}
\right]
\; {\bf \tau}_f \; \Phi
 =   \bar{\psi} \; \gamma^0 \; U^\dag
\left[
\begin{array}{ccc}
\Gamma^0 \Gamma^a & 0  \\
0 & \Gamma^0 \Gamma^a
\end{array}
\right] {\bf \tau}_f \;
U \; \psi \,.
\end{equation}
%
The complete list of mapping for quark bilinears is
given in Table \ref{dictionary}.

%
\begin{table}[thb]
\caption{Transformation between quark bilinears in $1+1$ 
and $3+1$ dimensions.}
\label{dictionary}
\begin{center}
\vspace{0.2cm}
\begin{tabular}{|c|c|c|c|c|}
\hline
&\hspace{0.5cm} ${\bf 1}$ \hspace{0.5cm} 
&\hspace{0.5cm} $\Gamma^5$\hspace{0.5cm} 
&\hspace{0.5cm} $\Gamma^0$\hspace{0.5cm} 
&\hspace{0.5cm} $\Gamma^z$\hspace{0.5cm}  \\
\hline
${\bf 1}$ & ${\bf 1}$ & $\gamma^0 \gamma^z$ & $\gamma^0$ & $\gamma^z$
\\
${\bf \tau}_1$ 
& $ - \gamma^5 \gamma^1$ 
& $ - i \gamma^2$ 
& $ \gamma^5 \gamma^0 \gamma^1$ 
& $ - i \gamma^0 \gamma^2$ 
\\ 
${\bf \tau}_2$
& $ - \gamma^5 \gamma^2$ 
& $ i \gamma^1 $
& $  \gamma^5 \gamma^0 \gamma^2$
& $ i \gamma^0 \gamma^1$
\\
${\bf \tau}_3$
& $ \gamma^5 \gamma^0 \gamma^z$
& $\gamma^5$
& $- \gamma^5 \gamma^z$
& $- \gamma^5 \gamma^0$ 
\\
\hline
\end{tabular}
\end{center}
\end{table}

We can also use these results to construct the relevant effective
Lagrangian for a model in which the gluon propagator is
a sum of a Gribov-Zwanziger term, Eq. (\ref{gluon_propagator}),
and a perturbative term.  Integration over the former gives QCD
in two dimensions, while integration over the perturbative gluon 
propagator give a non-Abelian Thirring model \cite{son}.  Thus the
general effective model is a gauged, non-Abelian Thirring model.
As noted in the Introduction, we consider only a region of intermediate
$\mu$, where the effects of the Thirring model can be neglected.

\section{Mapping a Fermi Sea of Massless Quarks onto the Vacuum}
\label{2D_ferm}

In two spacetime dimensions, for massless quarks
one can directly map the theory at $\mu \neq 0$ onto that
in vacuum, $\mu = 0$.  This has been noted before, in various guises,
in the literature before \cite{fischler}, especially by 
Christiansen and Schaposnik \cite{fischler}.  Hopefully our discussion
adds clarity.

Consider the following transformation of the quark fields:
\begin{equation}
\Phi = \exp\left( - i\, \mu \, z \, \Gamma^5 \right) \Phi' \; .
\label{anom_transf}
\end{equation}
This transformation is defined to be the same for each of the two
dimensional ``flavors'', $\varphi_\uparrow$ and $\varphi_\downarrow$, and so preserves
the flavor symmetry.  Under this transformation, the Lagrangian becomes
\begin{equation}
{\cal L}^{{\rm 2D}}_{{\rm eff}} 
= \overline{\Phi} \big[
 i \, \Gamma^\mu \, (\partial_\mu + i g_{{\rm 2D}} \, A_\mu) 
 +\mu \, \Gamma^0 \big] \Phi
=  \overline{\Phi}' \big[
 i \, \Gamma^\mu \, (\partial_\mu + i g_{{\rm 2D}}\, A_\mu)
 \big] \Phi'.
\label{trans_massless_qks}
\end{equation}
That is, by redefining the quark fields, we have {\it completely}
eliminated the chemical potential; one has transformed the theory
from one in the presence of a Fermi sea to that in vacuum.
This happens because when $\Gamma^z \partial_z$ acts upon 
$\exp(i \mu \, z \, \Gamma^5 )$, it equals $\mu \Gamma^z \Gamma^5$,
which by Eq. (\ref{twodim_identity}), equals $- \mu \Gamma^0$, and
so cancels the term for the quark chemical potential in the original
Lagrangian.

As we discuss in Sec. \ref{massive_quarks},
this is special to massless quarks, and does
not hold for massive quarks.  It also does not hold in higher dimensions,
where the effects of transverse fluctuations obviate any such correspondence.

We then have a quandry: there is a nonzero density of quarks in the
original theory, with $\mu \neq 0$.  The vacuum has no such density, so
where did it go?  The answer is that the transformation of 
Eq. (\ref{anom_transf}) is anomalous, involving the Dirac matrix
$\Gamma^5$.  

One can show that the correct quark density is given, precisely,
by the anomaly.  There are many ways of doing the calculation; in 
Appendix \ref{appendixA}
we give the computation based upon operator regularization.
The computation also shows that the only quark bilinear to receive
an anomalous contribution is that for quark number.  For other operators,
the transformation from $\Phi$ to $\Phi'$ can be computed naively.

The most interesting transformation is for the chiral condensate.
Using Eq. (\ref{mass_transf}), we write the a chiral condensate for
$\Phi$, in terms of $\Phi\,'$:
\begin{equation}
\overline{\Phi}\,' \, \Phi\,'
= {\rm cos}(2\mu z) \; \overline{\Phi} \, \Phi
- i \, {\rm sin}(2\mu z) \; \overline{\Phi} \, \Gamma^5 \, \Phi \; .
\label{trans_mass}
\end{equation}
Assume that there is chiral symmetry breaking in vacuum,
so that $\langle \overline{\Phi}\,' \, \Phi\,' \rangle \neq 0$.
Actually, in $1+1$ dimensions fluctuations disorder the system,
and only leave quasi long range order \cite{tsvelik}.  
Since this is due to fluctuations,
at large $\Nc$ such disorder only occurs
over distances exponential in $\Nc$ \cite{wzw2,wzw3}.  Neglecting
such details, if chiral
symmetry breaking occurs in vacuum, then it
also occurs in the presence of a Fermi sea, in the following manner:
\begin{equation}
\langle\overline{\Phi} \, \Phi \rangle= 
{\rm cos}(2\mu z) \; \langle \overline{\Phi}\,' \, \Phi\,' \rangle
\;\;\; ; \;\;\;
\langle \overline{\Phi} \, \Gamma^5 \, \Phi \rangle =
- i \, {\rm sin}(2\mu z)
\; \langle \overline{\Phi}\,' \, \Phi\,' \rangle \,.
\;
\end{equation}
This is the strict analogy of Migdal's pion condensation
\cite{pion_condensation}
in $1+1$ dimensions.  Chiral symmetry is broken, but by a spiral in
the two possible directions, between
$\overline{\Phi}\Phi$ and $\overline{\Phi}\Gamma^5\Phi$.
Sch\"on and Thies \cite{thies} termed 
this as a ``chiral spiral'', and we adopt their evocative name,
and so refer to our result as a Quarkyonic Chiral Spiral.

We can also understand why exciton pairing is not favored.
An exciton condensate corresponds
to $\psi^\dag_{R+} \psi_{L+}$, which is a spin $1$ operator.
In the effective theory, the corresponding operators are
$\overline{\Phi} \Gamma^z \tau_{1,2}\Phi$ and
$\overline{\Phi} \Gamma^0 \tau_{1,2}\Phi$.
These operators are flavor non-singlet.  
In two dimensions in vacuum, though, it is expected that the spontaneous
breaking of chiral symmetry proceeds through 
condensates which are flavor singlets, and not through
condensates which carry flavor.

We conclude this section by noting that the extended flavor symmetry
in $1+1$ dimensions is special to the vector-like interactions
of QCD, and is not a generic property of theories in $3+1$ dimensions.
Consider, for example, a general Nambu-Jona-Lasino (NJL) model,
with interactions such as $(\overline{\psi} \psi)^2$.  This is
the square of a mass term, and so is not invariant under the 
anomalous chiral transformation of Eq. (\ref{anom_transf}).  
In accord with this,
Nickel found that in NJL models at nonzero density, the 
thermodynamically favored ground state is a crystal, but not a chiral
spiral \cite{nickel}.  

\section{Mapping Fluctuations for a Fermi Sea of Massive Quarks}
\label{massive_quarks}

In this section we show how for massive quarks, excitations near the
Fermi surface can be mapped onto a (modified) theory of the vacuum.
For massless quarks, this could be done for the entire theory; here,
it is only for fluctuations near the Fermi surface.

Let the quark mass be $m$, so the Fermi momentum is related to the
chemical potential as $\mu^2 = p_F^2 + m^2$.  We work in the extreme
Quarkyonic limit, where $\mu \gg \lqcd$.  Starting with the theory
in $3+1$ dimensions, by neglecting the transverse momenta we
obtain an effective theory in $1+1$ dimensions,
\begin{equation}
{\cal L}^{{\rm 2d}}_{{\rm eff}} 
= \overline{\Phi} \big[
i \, \Gamma^\mu  \left(\partial_\mu + i \, g_{{\rm 2d}} A_\mu\right) 
+ \mu \, \Gamma^0 -m\big] \Phi \,.
\end{equation}

In this case, we could perform the transformation of Eq. (\ref{anom_transf}),
and so eliminate the term $\sim \mu \Gamma^0$ from the action, as in
Eq. (\ref{trans_massless_qks}).  The chemical potential is still in
the action, though, through the transformation of the mass term
into a complicated, position dependent ``mass'', 
as in Eq. (\ref{trans_mass}).

While one cannot make an exact correspondence to the vacuum theory,
one can still make an interesting, if more limited, correspondence.
Consider not all fluctuations in the theory, but just those near the Fermi 
surface.  As is well known \cite{tsvelik}, even massive particles
have a massless dispersion near the Fermi surface, with a modification
to the speed of light:
\begin{equation}
p_0 = \sqrt{(p_F + \delta p)^2 + m^2} - \mu \approx  v_{F} \, \delta \, p
\;\;\; ; \;\;\; v_F = \frac{|p_F|}{\mu} \; .
\label{massive_disp_rel}
\end{equation}
As for the massless case, we can neglect anti-quarks, deep in the Fermi sea.

First, let us decompose the spectrum of quarks.
We have two Fermi surfaces, $\pm |p_{F}|$, and 
have particle and hole excitations at each Fermi sea,
with energy $E\simeq \pm v_{F}|\delta p|$.
We identify the excitation around the Fermi sea with $+|p_{F}|$ 
as a right moving fermion,
and with $-|p_{F}|$ as a left moving fermion.
The effective Lagrangian becomes 
\begin{equation}
{ \cal L}^{\rm 2d,massive}_{\rm eff} 
= \overline{\Phi} \big[
i \, \Gamma^0  \left(\partial_0 + i A_0\right) 
+i \, v_F \, \Gamma^z  \partial_z
\big] \Phi \, ,
\label{massive_eff_lag}
\end{equation}
where we have assumed $A_z = 0$ gauge.  This is the same theory as
for the massless case, Eq. (\ref{trans_massless_qks}), except 
the speed of light is not one, but $v_F$.

These elementary manipulations can be used to explain the nature of
chiral spirals in exactly soluble models in $1+1$ dimensions
\cite{thies,bringoltz}.

For massless quarks, one maps the complete theory, in the presence of
a Fermi sea, onto the vacuum.  Thus the computation of 
$\langle \overline{\Phi}\,' \Phi\,' \rangle$ is complete, and 
both $\langle \overline{\Phi} \Phi \rangle$
and $\langle \overline{\Phi} \Gamma^5 \Phi \rangle$ 
oscillate about zero.  There
are several examples, such as the Gross-Neveu model for
a large number of flavors, in which this can be computed analytically
\cite{thies}.

Bringoltz considered a nonzero density of massive quarks
for QCD in $1+1$ dimensions \cite{bringoltz}.
By numerical analysis of the theory in the canonical ensemble,
he showed that chiral spirals also arise for massive quarks.
These are due to quark and quark hole excitations about the
edge of the Fermi surface, Eq. (\ref{massive_eff_lag}).
As the density increases, $v_F \rightarrow 1$, and the massive
theory approaches the massless limit.

%
\section{Effective Theories for Excitations about the Fermi Surface}
\label{WZW}
The massless excitations near the Fermi surface can be 
described in terms of a Wess-Zumino-Novikov-Witten (WZNW) theory
\cite{wzw1,wzw2,wzw3}.
The dictionary between bosonic and fermionic currents for
color, flavor, and (baryon) charge elements is
\begin{align}
J_+^A = &\frac{i}{2\pi} \tr [h^{-1} (\partial_+ h) t_A]
 = \np \Psi_+^\dag t_A \Psi_+ \pn,\ 
J_-^A = \frac{i}{2\pi} \tr[h (\partial_- h^{-1}) t_A]
 = \np \Psi_-^\dag t_A \Psi_- \pn, \nonumber \\
J_+^f = &\frac{i}{2\pi} \tr[g^{-1} (\partial_+ g) \tau_f]
 = \np \Psi_+^\dag \tau_f \Psi_+ \pn,\  
J_-^f = \frac{i}{2\pi} \tr[g (\partial_- g^{-1}) \tau_f]
 = \np \Psi_-^\dag \tau_f \Psi_- \pn, \nonumber \\
J_+ = &\sqrt{\frac{\Nc \Nf}{2\pi}} \partial_+ \phi
 = \np \Psi_+^\dag \Psi_+ \pn,\  
J_- = \sqrt{\frac{\Nc \Nf}{2\pi}} \partial_- \phi
 = \np \Psi_-^\dag \Psi_- \pn \; ;
\end{align}
$t_A$ is the color matrix, and $h$ an element of $SU(\Nc)$;
$\tau_f$ is the flavor matrix, and $g$ an element of
$SU(2 \Nf)$.  Normal ordering of a composite operator $A$ is
denoted by $\np A \pn$.

For free Dirac fermions with 
$\Nc$ colors and $2\Nf$ flavors,
\begin{equation}
S = \int d^2x [\Psi_+ i\partial_- \Psi_+ + \Psi_- i \partial_+ \Psi_-] \,,
\end{equation}
the bosonized version is 
\begin{equation}
S = S_{U(1)} [\phi] + S_{2 \Nf}^\text{color}[h] + S_{\Nc}^\text{flavor}[g]
\label{wzw_action}
\end{equation}
with
\begin{align}
S_{U(1)}[\phi]
 =& \Nc \Nf \int d^2 x \; (\partial_\mu \phi)^2, \nonumber \\
S_{k}[l] 
 =& k\ \tr \bigg[ 
\frac{1}{16\pi} 
 \int d^2x \; \partial_\mu l \partial^\mu l^{-1} 
+ \frac{1}{24\pi} 
 \int d^3x \; \epsilon^{\mu \nu \lambda} 
 (l^{-1} \partial_\mu l) (l^{-1}\partial_\nu l) (l^{-1} \partial_\lambda l) 
\bigg]. \nonumber \\
\end{align}
This is a sum of a free massless scalar, for the $U(1)$ of baryon number,
a $SU(\Nc)$ WZNW model with level $2 \Nf$, and a $SU(2 \Nf)$ 
WZNW model with level $\Nc$.

A similar form can be derived for the theory of Dirac fermions coupled
to a gauge field \cite{wzw2,wzw3}.  The flavor part of the action,
$S_{\Nc}^\text{flavor}[g]$, is completely unchanged, because the currents
which define the flavor matrix $g$ are color singlets.  The color
dependent part of the action, $S_{2 \Nf}^\text{color}[h]$, becomes
that of a gauged WZNW action.  

The spectrum of a gauged WZNW model is involved \cite{wzw2,wzw3}.  However,
what we are most concerned about are excitations near the Fermi surface;
namely, are there gapless modes.  In this context, what is of greatest
concern is that is that the massless correlations of the {\it flavor}
WZNW action, $S_{\Nc}^\text{flavor}[g]$, dominate correlations over
large distances.  

Note that this is true for {\it both} massless and massive quarks.  In
either case, there are numerous gapless modes about the Fermi surface.  

\section{Conclusions}
\label{conclusions}

In this paper we approximate the confining potential
in a Quarkyonic phase as in 
Eq. (\ref{gluon_propagator}).  We then
find that Quarkyonic Chiral Spirals (QCS's) arise naturally,
and can be expressed in terms of an effective model in $1+1$ dimensions.
Our analysis is valid in the extreme
Quarkyonic limit, where $\mu \gg \lqcd$, within a narrow
skin of the surface of the Fermi sea, $\sim \lqcd$.

In this work we only considered the formation of a single chiral density
wave, in a fixed direction.  It is most likely, though, that the
entire Fermi surface is covered with patches of 
chiral density waves, in different directions \cite{son,other_cdw}.
The detailed manner in which the Fermi surface is covered with such
patches will be discussed separately \cite{tocome}.

Quarkyonic Chiral Spirals 
are reminiscent of the pion
condensates of Migdal \cite{pion_condensation}.  
Pion condensates arise in effective models of nucleons interacting with
pions: they are chiral spirals which oscillate as
$\langle \overline{\psi} \exp(2 i c f_\pi  z \gamma_5) \psi \rangle$,
where $c$ is a pure number and $f_\pi$ is the pion decay constant.
In contrast, a QCS arises from the interactions of quarks and gluons;
it is a chiral spiral not in chirality,
$\gamma_5$, but in spin, $\gamma^0 \gamma^z$.

It is of great interest that we find a QCS for massive quarks. 
Kaplan and Nelson suggested that in dense matter,
effective models of nucleons and kaons indicate that
there is a condensate for the $K^-$ field \cite{kaon_condensation}.  
This is constant in space, and so is unlike the chiral spiral which
we would expect in strange Quarkyonic matter.

For distances $x \gg 1/\mu$, QCS's have numerous modes with long ranged
correlations, from the correlations of the WZNW model, Sec. \ref{WZW}.  
It is possible that the long range correlations of the modes of the
WZNW model acquired finite range over distances $\sim 1/\lqcd$.  Even
so, this predicts {\it many} more light modes than expected otherwise.

The formation of QCS's has strong implications for the phase diagram of 
a gauge theory.  
The usual expectation is that one goes from a phase where
chiral symmetry breaking is broken through a constant condensate,
$\langle \overline{\psi} \psi \rangle \neq 0$, directly to a phase where it
vanishes, 
$\langle \overline{\psi} \psi \rangle = 0$.

If a QCS forms, chiral symmetry is broken, but through an order
parameter which differs from that in vacuum.  As for a pion
condensate, a QCS
spontaneously breaks both translational and rotational symmetries.
Thus there is a strict order parameter which differentiates between
ordinary chiral symmetry breaking, with a constant chiral order parameter,
and a QCS.  It also implies that there are exactly massless Goldstone
bosons in a QCS, from the spontaneously
broken symmetries of translation and rotation.
Similarly, there is an order parameter which differentiates between
a QCS, and a phase which is chirally symmetric (at least for massless
quarks).  

This implies that at zero temperature, as $\mu$ increases there is first
a Fermi sea.  There is then a well defined phase transition from 
a phase with constant 
$\langle \overline{\psi} \psi \rangle \neq 0$ to a QCS.  There is
then a second phase transition, from the QCS to a chirally symmetry
phase, with 
$\langle \overline{\psi} \psi \rangle = 0$.
This differs from models which exclude spatially dependent
condensates, Ref. \cite{glozman,szczepaniak}, but
similar to models which allow for then, Ref. \cite{nickel}.

Thus at $T = 0$ and $\mu \neq 0$, 
we predict the existence of an intermediate phase, with a QCS.
In Appendix \ref{appendixB} we show that quarks affect gluons at
asymptotically large $\mu \sim \Nc^{1/2}$.  At such large $\mu$,
deconfinement and chiral symmetry breaking surely occur.  Thus
suggests that at least for large $\Nc$, the region with a QCS is
also large.  Further, it is very possible that the phase transition
from a QCS, to a chirally symmetry phase, occurs before the theory
deconfines.

We conclude by suggesting that dynamical quarks do {\it not} easily
wash out a Quarkyonic phase, nor related effects, such as
Quarkyonic chiral spirals.
In QCD, it is known that the effects of screening, from dynamical quarks,
are not strong.  Notably, the
linear term in the quark anti-quark potential persists to rather
short distances, $\sim 0.2$~fm.  This distance is comparable to the
short range repulsion experienced by nucleons.  Short distances then
translates into high densities.  
It would be very interesting to estimate this effect within effective
models, such as approximate solutions to the 
Schwinger-Dyson equations.  

Our analysis clearly raises more questions than it answers.  However,
we hope that we have provided a different way for thinking about
cold, dense quark matter, and about the surprises which it might provide.
In the end, this is of direct relevance to neutron (Quarkyonic?) stars,
for which a wealth of experimental data will be forthcoming in the next
few years.

\section*{Acknowledgments}

The research of R. D. Pisarski and L. McLerran is supported under 
DOE Contract No. DE-AC02-98CH10886.  R. D. Pisarski also thanks the
Alexander von Humboldt Foundation for their support.
This research of Y. Hidaka is supported by the Grant-in-Aid for
the Global COE Program ``The Next Generation of Physics, 
Spun from Universality and Emergence'' from the Ministry of 
Education, Culture, Sports, Science and Technology (MEXT) of Japan.
T. Kojo is supported by Special Posdoctoral Research Program of RIKEN.
L. McLerran gratefully acknowledges conversation with Thomas Sch\"afer, who 
insisted that chiral symmetry
breaking must occur in the form of chiral density waves at large $\Nc$.  
We also thank Gokce Basar, Barak Bringoltz,
Michael Buchoff, Aleksey Cherman, Thomas Cohen, 
Gerald Dunne, Robert Konik, Alex Kovner, Dominik Nickel, and
Alexei Tsvelik for useful discussions and comments.


\appendix

\section{Screening by Dynamical Quarks at Large $\Nc$}
\label{appendixB}

In this appendix we show that at a nonzero temperature, $T$,
screening by dynamical quarks enters
when $\mu \sim \Nc^{1/2} T$. 

Consider, for example, the 
square of Debye screening mass at one loop order \cite{mclerran}.  Gluons
contribute $\sim g^2 \Nc T^2$, quarks give $\sim g^2 \Nf \mu^2$.
Balancing the two terms, quarks start affecting gluons when
when $\mu \sim \Nc^{1/2}T$.

Next consider the same analysis for the free energy.
Contributions from gluons are $\sim \Nc^2 \, T^4$,
while those from quarks are $\sim \Nc \Nf$ times
powers of $\mu$ and $T$: $\mu^4$, $\mu^2 \, T^2$, and $T^4$, see
Eq. (\ref{semi3}) below.   Of course at $\sim g^4$ and beyond, the
quark and gluon contributions get mixed up together, but this does not
affect our power counting in $\Nc$.  
If we then balance the leading term for
gluons, $\sim \Nc^2 \, T^4$, against that of quarks,
$\sim \Nc \Nf \mu^4$, we estimate that quarks dominate when
$\mu \sim \Nc^{1/4}T$ \cite{mclerran}, and not $\sim \Nc^{1/2}T$, as for
the Debye mass.

With the free energy, however, one must take more care.  
One cannot simply equate the magnitude fo the free energies,
but remember that deconfinement is only defined
by a change in the relevant order parameter, which is the (renormalized)
Polyakov loop \cite{semi}.  
This is generated by a nontrivial distribution in the
eigenvalues of the thermal Wilson line.
This can be modeled by expanding about a
constant expectation for the timelike component of the vector potential,
\begin{equation}
A_0 = \frac{Q}{g} \; .
\label{semi1}
\end{equation}
$Q$, is a diagonal matrix in color space, and is traceless, as 
a sum over elements of $SU(\Nc)$.
$Q$ has dimensions of mass, where the mass scale is set by the temperature,
$T$.  We do not
need to know the explicit distribution of the $Q$'s in order to estimate
how large $\mu$ must be for quarks to affect the $Q$-distribution.

At very large $\mu$ 
we can compute the quark determinant, in the presence of this background
field, to one loop order.  A constant field for
$A_0$ acts like an imaginary chemical potential for color,
and so it is natural to introduce a color dependent
chemical potential \cite{altes},
\begin{equation}
\widetilde{\mu} = \mu + i \, Q \; .
\label{semi2}
\end{equation}
At one loop order
one obtains the usual result for the pressure, with $\mu$ replaced by
$\widetilde{\mu}$:
\begin{equation}
p_\text{quark} = \Nf \left( \frac{1}{12 \pi^2} \; {\rm tr} \; \widetilde{\mu}^4
+ \frac{T^2}{6} \; {\rm tr} \; \widetilde{\mu}^2 + \frac{7 \pi^2}{180}\;
\Nc \; T^4 \right) \; .
\label{semi3}
\end{equation}
For large $\mu$,
the dominant contribution is from the expansion of the first term,
$\sim \Nf \, {\rm tr} \, \widetilde{\mu}^4\sim \mu^4$, as we estimated
above.  While this is as large as the gluon contribution when
$\mu \sim \Nc^{1/4}T$, that does not matter, 
since this term is {\it independent} of
$Q$.  This holds order by order in perturbation theory, simply
because $Q$ has dimensions of mass.
The next term is from
the expansion of $\sim \Nf \, {\rm tr} \, \widetilde{\mu}^4$,
equal to $\sim \mu^3 \, {\rm tr} \, Q$; this vanishes, though,
because $Q$ is a traceless matrix.

The leading term which is $Q$-dependent is the next term in the
expansion of ${\rm tr} \, \widetilde{\mu}^4$, which is
$\sim \mu^2 \, {\rm tr} \, Q^2$.
Since the trace is $\sim \Nc$, this is as large as the gluon
contribution, $\sim \Nc^2$, when $\mu \sim \Nc^{1/2}T$.
This agrees with our estimate from the Debye mass.

At large $\Nc$, since the quarks do not affect gluons
until $\mu \sim \Nc^{1/2}T$, the boundary from the confining,
to the deconfining, phase is a straight line in the plane
of $T$ and $\mu$ \cite{mclerran}.  
The pure glue theory has a global symmetry of $Z(\Nc)$, which for
large $\Nc$ is approximately $U(1)$.  
At large $\mu$, the leading term from quarks is $\sim {\rm tr} \, Q^2$,
and breaks the $Z(\Nc)$ symmetry.  Like other terms from quarks,
this favors a real expectation value for the Polyakov loop.
Such a term acts to wash out the line of first order transitions.

There are two possibilities.  One is that the quarks produce a critical
endpoint for deconfinement.  The other is that the first line for deconfinement
bends and meets the axis for $T=0$.  Since all of our analysis depends
upon $T \neq 0$, we favor the former.

\section{Anomalous Baryon Number}
\label{appendixA}

In this appendix we compute the baryon number generated by
the anomalous transformation in Eq. (\ref{anom_transf}).
Consider the operator for baryon number, computed with
point splitting:
\begin{equation}
\begin{split}
\lim_{\epsilon \to 0}
\langle \overline{\Phi}(x+\epsilon) \Gamma^0 \Phi(x) 
\rangle_{\mu \neq 0}
=& 
\lim_{\epsilon \to 0}
{{
 \int {\cal D}\overline{ \Phi} {\cal D} \Phi 
\ \overline{\Phi}(x+\epsilon) \Gamma^0 \Phi(x)
e^{iS[\Phi;\mu\neq 0]} 
} \over {
  Z[\mu \neq 0] 
}} 
 \\
=& 
\lim_{\epsilon \to 0} 
{{
\int {\cal D}\overline{ \Phi}' {\cal D} \Phi' 
\ \overline{\Phi}'(x+\epsilon) 
e^{i\mu \epsilon_z \Gamma^5}
\Gamma^0 \Phi'(x)
e^{iS[\Phi';\mu = 0]}
} \over { 
Z[\mu=0]
}}
  \\
=& 
2 \; \lim_{\epsilon \to 0}
\bigg(\frac{i}{2\pi \epsilon^2} \bigg)
\bigg[
 \tr[\Gamma^0 \Slash{\epsilon} ]
+ i\mu \epsilon_z \tr
[\Gamma^5\Gamma^0 \Slash{\epsilon}] 
+ O(\epsilon^3)
\bigg] \\
=& 2 \; \lim_{\epsilon \to 0} 
\bigg[ \frac{i \epsilon_0}{\pi \epsilon^2}
- \frac{\mu \epsilon_z^2 }{\pi \epsilon^2} \bigg] \,.
\label{NB}
\end{split}
\end{equation}
We take the symmetric limit to preserve Lorentz symmetry in the
ultraviolet regime,
\begin{equation}
\lim_{\epsilon \to 0} \frac{ \epsilon_{\mu}\epsilon_{\nu} }{\epsilon^2} 
= \frac{g_{\mu \nu} }{2} \,.
\end{equation}
The first term is odd in $\sim \epsilon_0$, and so vanishes after
averaging over both directions.  Hence
\begin{equation}
\lim_{\epsilon \to 0}
\langle \overline{\Phi}(x+\epsilon) \Gamma^0 \Phi(x) 
\rangle_{\mu \neq 0}
= \frac{\mu}{\pi} \,.
\end{equation}

This equals the baryon number density for a gas of two flavors of free quarks,
with chemical potential $\mu$.

The computation can be repeated for other operators, such as
$\langle \overline{\Phi} \Phi \rangle$,
$\langle \overline{\Phi} \Gamma^5 \Phi \rangle$,
and $\langle \overline{\Phi} \Gamma^z\Phi \rangle$.  These
operators do not have anomalous terms, since
\begin{equation}
\tr [\Gamma^5\Gamma^5 \Slash{\epsilon}]
= 0, \ \ 
\tr [\Gamma^5\Slash{\epsilon}]
= 0,\ \ 
\frac{\epsilon_z}{\epsilon^2} 
\tr [\Gamma^5\Gamma^z \Slash{\epsilon}]
= \frac{\epsilon_z \epsilon_0}{\epsilon^2} \rightarrow 0 \; ,
\end{equation}
respectively. For these condensates, the explicit phase
dependence, in transforming from $\Phi$ to $\Phi'$, must
be taken into account, but there are no additional terms from the
anomaly.

\end{document}